\documentclass[aps,twocolumn,showpacs,amsmath,amssymb,superscriptaddress,pra,nofootinbib,10pt]{revtex4-1}
\usepackage{graphicx}
\usepackage{amsmath}
\usepackage{textgreek}
\usepackage{etoolbox}
\usepackage[hidelinks]{hyperref}

\usepackage[top = 2.0cm, bottom = 2.00cm, left = 2.00cm, right = 2.0cm]{geometry}

\makeatletter
\pretocmd{\chapter}{\addtocontents{toc}{\protect\addvspace{10\p@}}}{}{}
\pretocmd{\section}{\addtocontents{toc}{\protect\addvspace{0\p@}}}{}{}
\makeatother

\usepackage{array}
\newcolumntype{R}[1]{>{\raggedright\let\newline\\\arraybackslash\hspace{0pt}}m{#1}}
\newcolumntype{C}[1]{>{\centering\let\newline\\\arraybackslash\hspace{0pt}}m{#1}}
\newcolumntype{L}[1]{>{\raggedleft\let\newline\\\arraybackslash\hspace{0pt}}m{#1}}
\usepackage{hhline}

\usepackage{setspace}

\begin{document}

\title{Cryogenic platform for coupling color centers in diamond \\membranes to a fiberbased microcavity}

\author{M.~Salz}\email{m.salz@uni-mainz.de}
\affiliation{QUANTUM, Institut f\"ur Physik, Johannes Gutenberg-Universit\"at Mainz, Staudingerweg 7, D-55128 Mainz, Germany}
\author{Y.~Herrmann}
\affiliation{QUANTUM, Institut f\"ur Physik, Johannes Gutenberg-Universit\"at Mainz, Staudingerweg 7, D-55128 Mainz, Germany}
\author{A.~Nadarajah}
\affiliation{School of Physics, University of Melbourne, Parkville VIC 3010, Australia}
\author{A.~Stahl}
\affiliation{QUANTUM, Institut f\"ur Physik, Johannes Gutenberg-Universit\"at Mainz, Staudingerweg 7, D-55128 Mainz, Germany}
\author{M.~Hettrich}
\affiliation{Alpine Quantum Technologies GmbH, Technikerstrasse 17 / 1, A-6020 Innsbruck, Austria}
\author{A.~Stacey}
\affiliation{School of Physics, University of Melbourne, Parkville VIC 3010, Australia}
\author{S.~Prawer}
\affiliation{School of Physics, University of Melbourne, Parkville VIC 3010, Australia}
\author{D.~Hunger}
\affiliation{Physikalisches Institut, Karlsruher Institut f\"ur Technologie, Wolfgang-Gaede-Straße 1, D-76131 Karlsruhe, Germany}
\author{F.~Schmidt-Kaler}
\affiliation{QUANTUM, Institut f\"ur Physik, Johannes Gutenberg-Universit\"at Mainz, Staudingerweg 7, D-55128 Mainz, Germany}

\date{\today}

\begin{abstract}
	We operate a fiberbased cavity with an inserted diamond membrane containing ensembles of silicon vacancy centers (SiV$^-$) at cryogenic temperatures $ \geq4~$K. The setup, sample fabrication and spectroscopic characterization is described, together with a demonstration of the cavity influence by the Purcell effect. This paves the way towards solid state qubits coupled to optical interfaces as long-lived quantum memories.
\end{abstract}

\maketitle

\section{Introduction}\label{intro}

Color centers in diamond emerged as promising candidates for a broad field of applications, including quantum sensing~\cite{Fuchs2018}, quantum communication~\cite{Leifgen2014,Humphreys2018, Nguyen2019a, Bradley2019} and quantum memories~\cite{Maurer2012,Bar-Gill2013}. Such applications require stable solid state emitters with lifetime-limited emission lines, which for several color center species, can be achieved using a high-quality, low strain crystal host at cryogenic temperatures~\cite{Sukachev2017a,Iwasaki2015,Trusheim2020}.\\
Various approaches aim to improve the photon collection from solid state emitters, employing solid immersion lenses~\cite{Robledo2011,Hadden2010}, nanopillars~\cite{Maletinsky2012} or waveguides~\cite{Mouradian2015}. Coupling the emitter to a nanophotonic~\cite{Zhang2018,Nguyen2019a,Riedrich-Moeller2014}, nano-fiberbased~\cite{Romagnoli2020} or open \cite{Albrecht2013,Riedel2017,Kaupp2016,Haeussler2019,Nair2019} optical cavity can be used to both enhance the emission into the zero-phonon line (ZPL) for emitter species with small Debye-Waller factors, like the NV$ ^{-} $ center, as well as to funnel the emission into a well-collectable optical mode via the Purcell effect. While photonic crystal cavities are attractive due to strong mode confinement~\cite{Zhang2018} and their intrinsic robustness, processing the crystal environment can lead to spectral diffusion and the optical outcoupling of the signal is challenging~\cite{Nguyen2019b}. Open, fiberbased Fabry-P\'{e}rot microcavities have been successfully utilized for various physical systems, like neutral atoms~\cite{Colombe2007,Gallego2018}, ions~\cite{Brandstaetter2013,Meyer2015,Pfister2016} or quantum dots~\cite{Muller2009}, as they allow for direct coupling between the fiber and the cavity mode, small mode volumes as well as full spectral and spatial tunability. In recent experiments, color centers incorporated in nanodiamonds were coupled to fiberbased microcavities~\cite{Albrecht2013,Kaupp2016,Benedikter2017}. However, the embedded emitters often suffer from spectral diffusion~\cite{Lindner2018} and insufficient photostability~\cite{Benedikter2017}. Recently, micrometer thin diamond membranes~\cite{Ruf2019} have shown to be promising hosts for color centers used in microcavity experiments~\cite{Riedel2017,Haeussler2019,Nair2019}.\\
Here, we present an experimental platform to couple color centers incorporated in single crystal diamond (SCD) membranes to a fiberbased microcavity at cryogenic temperatures. We integrate the SCD membranes with a thickness of $(1.42\,\pm\,0.02)~$\textmu m hosting ensembles of silicon vacancy centers (SiV$^{-}$) into the cavity and investigate their spectral properties down to 4~K. A Purcell enhancement of the excited state decay into the cavity mode is demonstrated.\\
This paper is organized as follows: We describe the experimental setup of the fiber-cavity and the integration of the diamond SCD membrane. Then we show spectroscopic measurements at temperatures between 300~K and 4~K and demonstrate the Purcell enhanced photon emission. 

\section{Experimental platform}
\label{sec:expPlatform}
\begin{figure}
	\centering
	\includegraphics[width=0.49\textwidth]{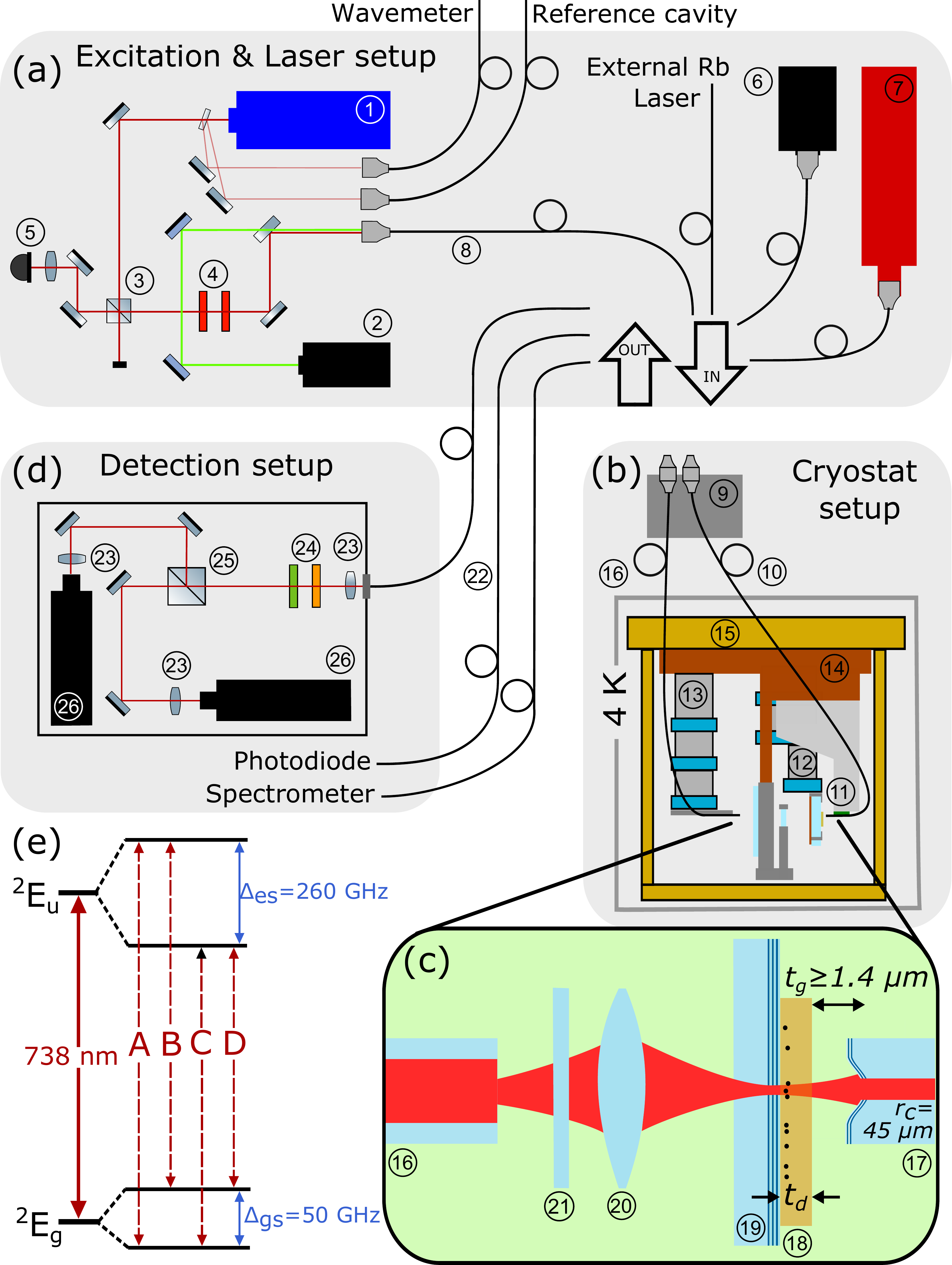}
	\caption{Experimental overview. \textbf{a} Excitation \& laser setup: (1) red diode laser, (2) green solid-state laser, (3) polarizing beam splitter, (4) polarization control, (5) photodiode, (6) pulsed green laser, (7) broadband light source, (8) PM transfer fiber. \textbf{b} Cryostat Setup: (9) fiber breakout box, (10) PM inside cryostat, (11) locking piezo, (12,13) 3D nanopositioner stack, (14) experimental insert baseplate, (15) cryostat baseplate, IVC, (16) MM inside cryostat. \textbf{c} Beam through experimental insert: (17) microcavity SM featuring a concave surface with a radius of curvature $r_c$ at the center, (18) diamond membrane with thickness $t_d$, (19) dielectric mirror, (20) aspheric lens, (21) longpass filter. $t_g$ denotes the gap between the fiber tip and the diamond membrane. \textbf{d} Detection setup: (22) MM transfer fiber, (23) collimation optics, (24) optical filters, (25) 50:50 non-polarizing beam splitter, (26) avalanche photodiode. \textbf{e} SiV$^-$ level scheme for vanishing magnetic field.}
	\label{fig:expSetup}
\end{figure}

All measurements are performed on a fully tunable fiberbased microcavity (complete setup in Fig.~\ref{fig:expSetup}). The cavity has a plano-concave Fabry-P\'{e}rot design, composed of a plane half inch mirror and a singlemode (SM) fiber with a dimple (45~\textmu m radius of curvature) inside a plateau (25~\textmu m in diameter) at the center of the fiber tip~\cite{Kaupp2016}. The dimple is produced by CO$ _2 $ laser ablation, which creates a Gaussian-shape profile with a surface roughness of typically $ \sigma_{\text{rms}}=0.2~$nm~\cite{Hunger2012}. The plane mirror and the fiber ends facet are coated with a dielectric stack that reflects 99.85\% and transmits 1480~ppm\footnote{parts per million} of the incident light at 736~nm. The cavity SM fiber is glued into a stainless steel cannula and mounted on a shear piezo for fine length control and stabilization. The fiber is spliced to a po\-lar\-iza\-tion-maintaining (PM) transfer fiber. Light transmitted by the plane mirror is collected by an aspheric lens and coupled into a multimode (MM) fiber after passing a longpass filter cutting at 650~nm to suppress fiber background emission (Fig.~\ref{fig:expSetup}c).

In general, the SM side is used as input and the MM side as output port until stated otherwise. The MM is mounted to a 3D nanopositioner stack\footnote{Attocube ANPx101, attocube ANPz102} with position readout for reproducible placement of the fiber into the focus of the lens (Fig.~\ref{fig:expSetup}d). For coarse microcavity length adjustment and lateral control, the macroscopic mirror is mounted on a second 3D nanopositioner stack\footnote{Attocube ANPx311, attocube ANPz51}~\cite{Casabone2018}. In that way, the plane mirror can be moved several mm away from the cavity fiber tip to avoid crashing the fiber into the mirror during cool down. The base plate of the experimental insert, carrying the microcavity and the nanopositioners is screwed to the base plate of the cryostat insert.\\
The setup including the cavity and positioning devices is incorporated into a wet $\rm ^{3}He$-$\rm ^{4}He$ dilution refrigerator\footnote{Oxford Kelvinox 100}, allowing base temperatures of down to 23~mK. However, we only operate the system as bath cryostat with liquid He since it was not necessary to go to lower temperatures due the spectroscopic properties at 4~K (see Sec.~\ref{sec:SivInMC}). All compressors needed for the operation of the cryostat are external, greatly reducing the vibrational noise in the system. Both sides of the cavity are accessed from outside the cryostat via optical fibers. The PM and MM fiber are guided to the four intermediate temperature stages of the cryostat and are glued with cryogenic varnish for thermal and mechanical anchoring. At the top, the fibers are fed through a press seal to the outside of the cryostat into a home-built fiber breakout box. It provides optical in/out coupling to both sides of the cavity via standard APC connectors for the excitation and SiV$^-$ fluorescence light.\\
The microcavity is probed with a grating-stabilized diode laser\footnote{Toptica DL pro design} at 737~nm for cavity characterization and resonant excitation of the SiV$^-$. Off-resonant excitation is achieved with a solid-state laser at 532 nm, which lies outside the stopband of the dielectric mirror coating, to enable the excitation of SiV$^-$ centers without the need of setting the cavity to be resonant to the laser frequency. To measure the microcavity length and dispersion relation for the incorporated membrane, we employ a compact broadband light source\footnote{Thorlabs SLS201L} coupled into the cavity. The excitation light is fiber coupled and connected to the breakout box at top of the cryostat. For detection, the output of the microcavity is either fiber-coupled to a photodiode, a grating spectrometer or an avalanche photodiode (APD) (Fig.~\ref{fig:expSetup}d). Spectral filters in front of the APD detector  result in a spectral detection window from 720~nm to 738~nm.

\subsection{Characterization of the empty cavity}

\begin{figure}
	\includegraphics[width=0.49\textwidth]{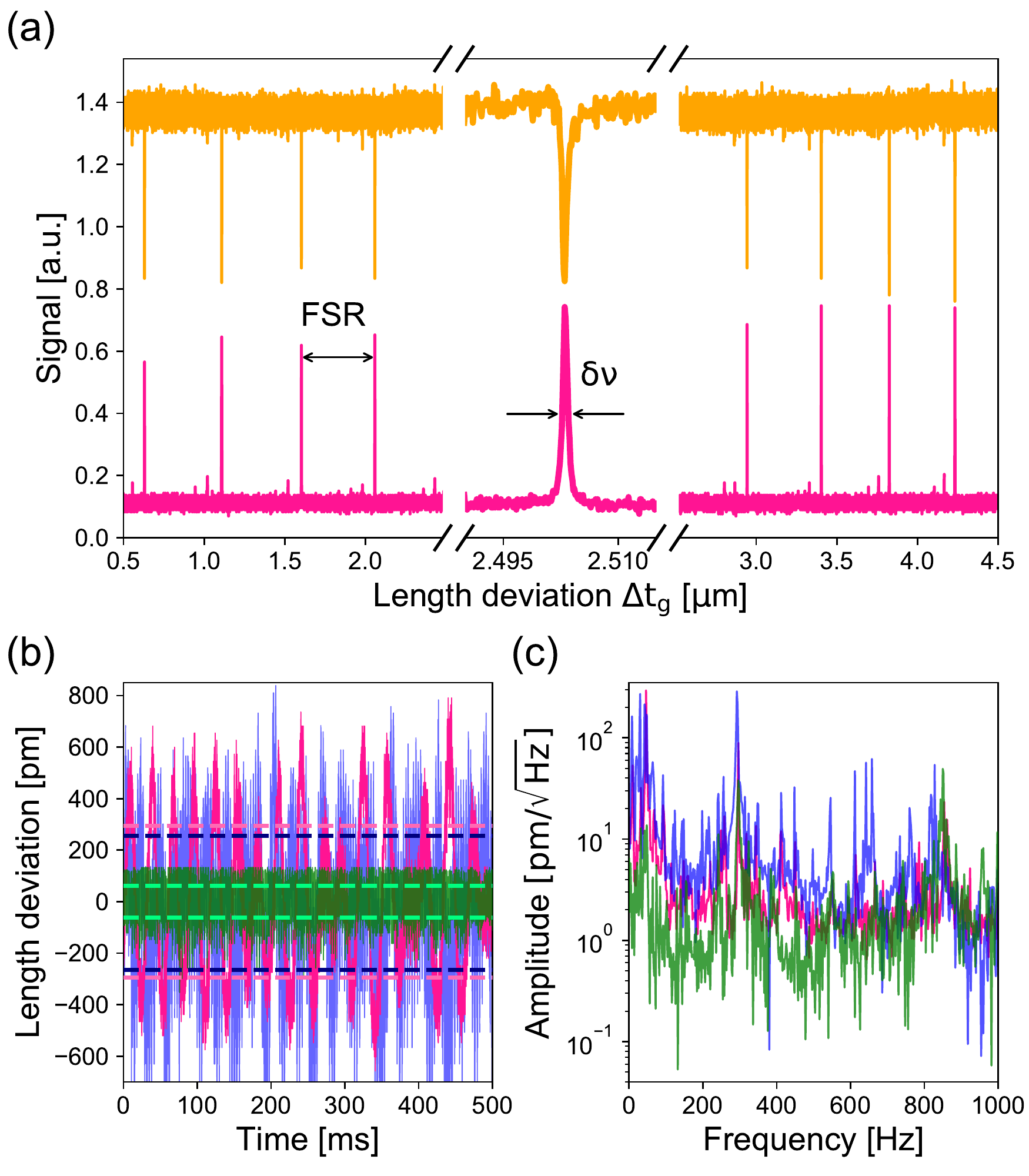}
	\caption{Performance of the bare microcavity under ambient and cryogenic conditions. \textbf{a} Scan of the microcavity around 15~\textmu m length showing reflection (yellow) and transmission signal (pink). The fundamental modes appear as highest peaks, separated by the free spectral range. The inset shows a single cavity resonance with full width at half maximum $\delta \nu$.\textbf{b} Locked (green) and unlocked (pink: 300~K, blue: 4~K) microcavity length deviation at a cavity length of 12~\textmu m, calculated from the transmission signal. The green (pink, blue) dashed lines indicate the the $\pm\,\sigma$ range of the locked (unlocked at 300~K, unlocked at 4~K) transmission. The side-of-fringe lock suppresses about 77\% of the length fluctuations. \textbf{c} Fourier spectrum of locked (green) and unlocked microcavity length fluctuation at 300~K (pink) and 4~K (blue), measured with the same cavity properties as in \textbf{b}. The dominant mechanical resonances below 50~Hz are compensated by the lock while higher frequencies are less suppressed.}
	\label{fig:bareCav}
\end{figure}

The microcavity shows stable operation in the length range from the radius of curvature of the concave fiber end, $r_c=(45\,\pm\,5)~$\textmu m, determined from interferometric measurements of the fiber profile, down to a minimum length of $ t_g=q\times\lambda/2=1.6$~\textmu m, limited by the depth of the concave profile of the fiber ends facet and the penetration of the light into the mirror stack. This corresponds to a fundamental mode order of $q\,=\,4$ with a mode waist of $w_0=1.4$~\textmu m, resulting in a mode volume of $V_{m}=5.8\,\lambda^3$ at 736~nm. We determine the cavity finesse  with laser light at 737~nm and scanning the microcavity length. A typical length scan with nine resonances in reflection and transmission can be seen in Fig.~\ref{fig:bareCav}a. For $t_g\leq10~$\textmu m, the finesse settles at $2200\,\pm\,100$, expected from the mirror coating. At longer cavity lengths, diffraction losses lead to a reduced finesse~\cite{Benedikter2015}. For $t_g$~=10~\textmu m, this finesse corresponds to a quality factor of $1.4\times 10^5$ for the empty cavity.\\
In order to actively stabilize the cavity length, we employ laser light at 780~nm, as this allows for spectral separation and simultaneous observation the SiV$^-$ fluorescence signal. A grating-stabilized diode laser locked on a Rb vapor cell resonance generates a side-of-fringe error signal from the cavity transmission. At 780~nm, the finesse is reduced to about 1000.\\
With the setup operated at 300~K and $t_g=(11.6\,\pm\,0.1)~$\textmu m, a reduction of the temporal length fluctuations from ($290\,\pm\,50$)~pm (150\%~cavity linewidth) to ($60\,\pm\,5$)~pm (30\% cavity linewidth) is achieved (Fig.~\ref{fig:bareCav}b). Fig.~\ref{fig:bareCav}c shows the Fourier spectrum of the cavity length deviation for the unlocked and locked case. The lock shows efficient suppression of noise up to frequencies of about 800~Hz. However, locking the cavity proves itself to be substantially harder at liquid He temperature (from ($ 260\,\pm\,30 $)~pm for the unlocked case down to ($ 90\,\pm\,20 $)~pm in the locked case at $ t_g=(34.2\,\pm\,0.1)~$\textmu m), mainly because of a ($3.4\, \pm\, 0.3$)-fold increase of vibrational noise at ($293\, \pm\, 1$)~Hz at 4~K compared to ($296\, \pm\, 1$)~Hz at 300~K. We conjecture eigenfrequencies of the nanopositioners, ranging at these lower frequencies, to be the reason for large vibration amplitudes.

\section{Diamond membrane}

In this chapter, we present a method to fabricate thin diamond membrane windows with high crystal quality and low surface roughness. This is required to introduce SiV$^-$ centers with favorable optical properties, like lifetime limited emission linewidths and long coherence times, into the microcavity, while maintaining a high cavity finesse. Furthermore, the position of the SiV$^-$ centers in the diamond needs to match the standing wave field of the cavity mode. This is achieved by the implantation of Si$^+$ atoms with well-defined energy into the crystal.
\subsection{Preparation and application}
\label{subSec:prepApp}

\begin{figure}
	\includegraphics[width=0.49\textwidth]{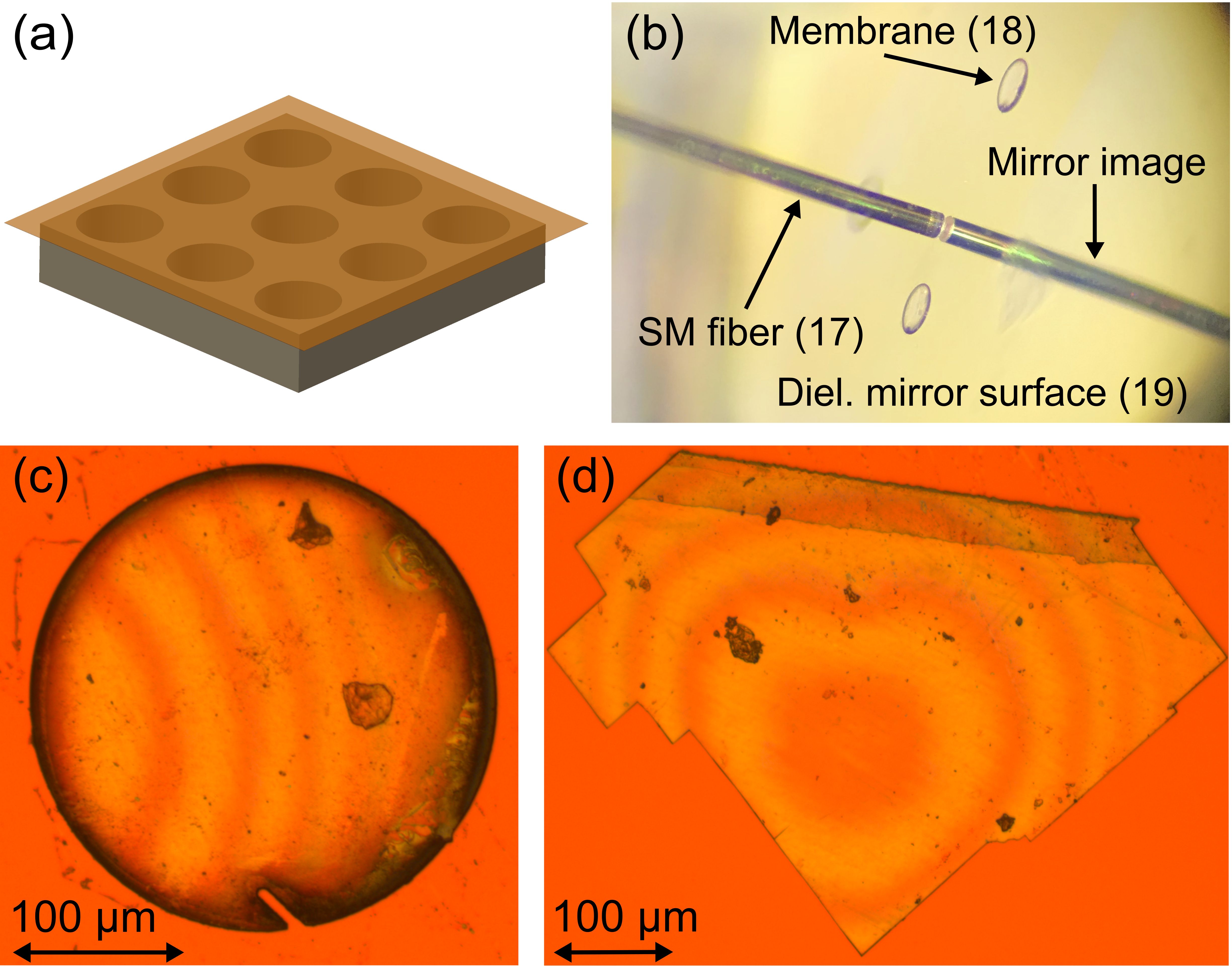}
	\caption{Diamond membrane samples on the plane mirror of the microcavity. \textbf{a} Sketch of the diamond membrane fused to a diamond scaffold with nine windows, before laser cutting. \textbf{b} Light microscope image of the cavity setup. The cavity fiber coming from the left side, together with the cavity mirror, forms the microcavity (fiber is mirrored at the right hand side). Diamond membranes are bonded on the cavity mirror. See Fig.~\ref{fig:expSetup}c for details. \textbf{c} Light microscope image of a cut out membrane window. The carbon debris, created from laser cutting, can been seen as the black material on the window edge. \textbf{d} Light microscope image of a broken membrane piece, applied to the mirror with tweezers. A layer of polycrystalline diamond underneath the upper edge is visible.}
	\label{fig:membraneApp}
\end{figure}

Fabricating the diamond membrane samples involves multiple steps. First, commercial high pressure, high temperature (HPHT) samples\footnote{Element Six, Type~1b,\\ (3.0~$\times$~3.0~$\times$~0.3)~mm} are used as seed substrates to begin the fabrication of the SCD membrane samples.\\
The substrates are implanted with high energy He ions (1~MeV, 5~$\times$~10$^{16}$~ions/cm$^2$) to create an amorphous layer $\approx$~1.7~\textmu m below the top diamond surface. The implanted samples are subsequently annealed at 1150$\,^\circ$C for 1~h in vacuum ($\approx$~5.0~$\times$~10$^{-6}$~torr) to convert the amorphous layer into a graphitic-like etchable layer. A single crystal homo-epitaxial growth was performed on the implanted samples to obtain a high-quality diamond overgrowth of $\approx$~6~\textmu m.\\
We then conduct a laser micromachining of a polycrystalline diamond to create supporting scaffolds with micro-channels for membranes, minimizing the risk of breakage and for easier handling. Then, we perform another growth process to fuse the scaffold into the over-grown sample along the contact and an electrochemical etching to lift-off the membrane from the substrate~\cite{Piracha2016,Piracha2016a}. Prior to ICP-RIE\footnote{inductively coupled plasma - reactive ion etching} processes, the membrane sample is mounted on a glass or Si substrate, such that the lifted-off side faces up. We first use an ArCl$_2$ based plasma to obtain the desired membrane thickness. We subsequently use an O$_2$ based plasma for a short duration ($\approx$~3~min) to remove any possible Ar and Cl contaminations~\cite{Appel2016}.\\
Si$^+$ implantation is carried out on the membrane windows using a microbeam at an energy of 110~keV with fluences from 4.3~$\times$~10$^8$ to 1.0~$\times$~10$^{14}$~ions/cm$^2$, containing 14 circular implantation regions in each window with a diameter of 45~\textmu m. The implantation energy is chosen such that the target depth of the Si$^+$ ions matches the first antinode of the intracavity light field (75 nm, simulated with SRIM\footnote{stopping and range of ions in matter}~\cite{Ziegler1985}). The implanted membrane sample is annealed at 1000$\,^\circ$C for 3~h in vacuum to promote the formation and activation of silicon~vacancies~\cite{Becker2018}. The sample is subsequently acid-cleaned and annealed at 500$\,^\circ$C in O$_2$ to remove any graphitic $sp^2$~layers~\cite{Tetienne2018,Fu2010}.\\
The SiV$^-$-implanted membrane windows (Fig.~\ref{fig:membraneApp}a) are laser cut and placed onto a clean Si~substrate. The dielectric mirror was cleaned using acetone, methanol, and IPA, followed by an additional 10~min O$_2$~plasma cleaning process, prior to apply membrane windows. Using a water drop on a thin wire attached to a needle, the membrane windows are drawn from Si~substrate and transferred to the dielectric mirror (Fig.~\ref{fig:membraneApp}b). However, some carbon debris at the edge of the diamond membrane windows can be seen (Fig.~\ref{fig:membraneApp}c). This debris leads in photoluminescence (PL) measurements (see Sec.~\ref{subSec:SiVSpec}) to a large, broadband background signal due to total internal reflection inside the membrane, even at positions without debris. Hence we investigate in the following only SiV$^{-}$ ensembles present in a broken membrane piece (Fig.~\ref{fig:membraneApp}d). For this sample, an unintended air gap between the membrane and the mirror is found which shifts the coupled modes of the microcavity (see Sec.~\ref{subSec:memCavSystem}).

\subsection{Integrated membrane-cavity system}
\label{subSec:memCavSystem}
\begin{figure}
	\includegraphics[width=0.49\textwidth]{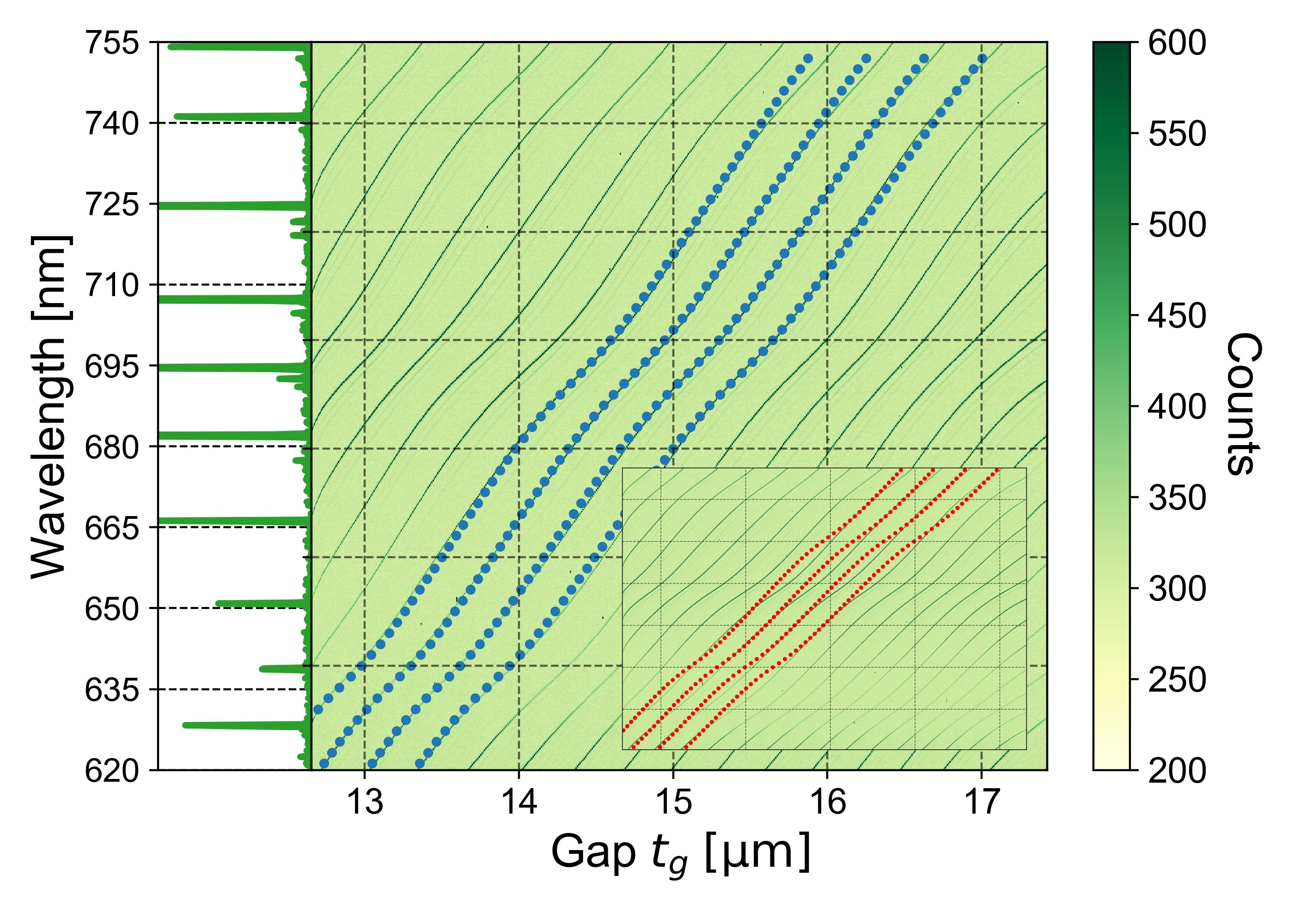}
	\caption{Transmission spectra of the cavity-membrane system for varying membrane-fiber gap. The left panel shows a single transmission spectrum for probing the system with white light. On the right, a series of length dependent transmission spectra is shown. The coupled modes are described by a transfer matrix model simulating all optical layers. Here an additional gap between membrane and macroscopic mirror of 250(50)~nm and a membrane thickness of 1.42(2)~\textmu m is assumed. The blue points are calculated from the matrix approach for the coupled fundamental orders of $q_{\text{gap}}\,$=\,42 to 45. The inset shows the matrix model simulation without a second gap.}
	\label{fig:dispRel}
\end{figure}

Introducing a thin layer of diamond into an optical resonator alters the cavity properties. First, both the surface roughness of the diamond membrane as well as dirt on this surface can lead to additional loss channels of the cavity, potentially reducing the finesse depending on the thickness of the membrane layer~\cite{vanDam2018}. The transmission of the mirror with the diamond layer attached can be significantly enlarged depending on the thickness of the diamond layer, as it can have a similar effect as an anti-reflective coating layer~\cite{Janitz2015}. This can drastically reduce the finesse as well as shift the probability of a photon which was emitted into the cavity mode leaving the cavity through the mentioned mirror. Additionally, the resonance conditions of the cavity are shifted.\\
Using an analytic 1D model~\cite{Janitz2015}, one finds that the cavity resonance frequencies are no longer linearly dependent on the cavity length but a complex dependency in the case of coupled \textit{air-} and \textit{diamond-like} modes. We use this model to calculate the diamond thickness and the gap between the diamond membrane and fiber mirror by fitting it to the measured resonance peaks. If we model only the independently determined diamond membrane thickness of ($ 1.42\, \pm\, 0.02)~$\textmu m, we find bad agreement with the measured data (see inset of Fig.~\ref{fig:dispRel}). However, if we use a transfer matrix model~\cite{Janitz2015}, describing both dielectric mirror stacks, sandwiched by a gap, the diamond membrane and a second gap of $ (250\,\pm\,50)~$nm between the plane mirror and the diamond membrane, we find excellent agreement of the calculated resonances with our measured data (Fig.~\ref{fig:dispRel}). The existence of a parasitic second gap is further confirmed from optical microscopy where we observe interference fringes and a non-perfect van der Waals bonding of the membranes. This is evidenced by a sudden shift of the membrane position when the cavity fiber was approached closely, presumably due to forces from electrostatic charging.
For the fiberbased cavity including the diamond membrane, the measured finesse depends on the exact position of the cavity mode on the diamond membrane. 
For the best case, a drop of the finesse down to 70\% of the bare cavity finesse is measured. The additional losses of $ (2100\, \pm \,600)~$ppm per round trip can be attributed to the diamond membrane. We conjecture the surface roughness of $\sigma_{\text{rms}}= (3.6\,\pm\,0.2)~$nm as main reason, which would explain additional scattering losses $ L_{\text{sc}} $ of up to 11700~ppm depending on the diamond thickness and hence the position of the diamond boundary with respect to the cavity mode standing wave field, estimated using an extended transfer matrix model with partially reflective rough interfaces~\cite{vanDam2018}.\\
We measure a maximum quality factor of $ (7.2\,\pm\,0.7)\times 10^4$ for $ L_{\text{eff}}=(20.0\,\pm\,0.5)~$\textmu m for the cavity including the diamond membrane.

\section{Coupling SiV$^-$ ensembles to the microcavity}\label{sec:SivInMC}

\subsection{Properties of SiV$^-$ ensembles in the diamond membrane}\label{subSec:SiVSpec}

\begin{figure}
	\includegraphics[width=0.49\textwidth]{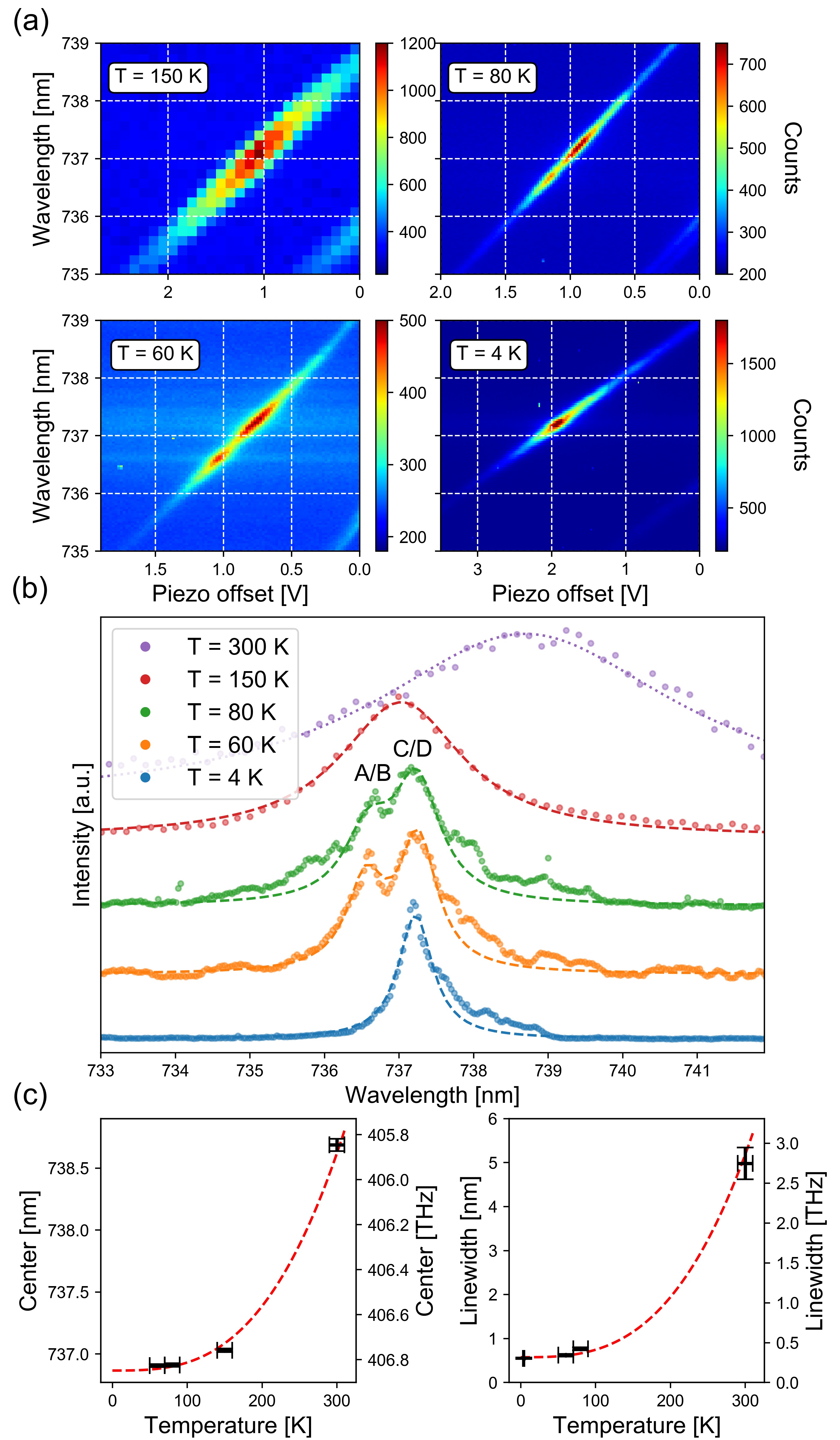}
	\caption{Fluorescence spectrum of the SiV$^-$ ensemble. \textbf{a} Cavity frequency scan over the SiV$^{-}$ resonance at different temperatures by scanning the nanopositioner piezo offset (increasing the voltage reduces the cavity length). At $T=150$~K, one resonance is observed. For $T=80$~K and $T=60$~K, two resonances can be resolved. At $T=4$~K, only the higher wavelength resonance remains. \textbf{b} Corresponding summed spectra for the scans shown in \textbf{a} (see text). The spectrum at 300~K is fitted by a single Lorentzian function (dotted line), while all other spectra are fitted by a double Lorentzian function (dashed lines). \textbf{c} The SiV$^-$ optical transition wavelength follows the expected cubic dependence on the temperature (red)~\cite{Jahnke2015} and settles around 736.9~nm for 4~K. Because only one peak remains at 4~K, no absolute position can be extracted here. The ZPL linewidth also shows the expected cubic dependence (red)~\cite{Jahnke2015}. Due to the vanishing doublet peak structure at 150~K, no linewidth value was extracted at this temperature.}
	\label{fig:linewidth}
\end{figure}

We perform PL spectroscopy on a SiV$^-$ ensemble with 532~nm light sent into the cavity via the SM fiber while scanning the cavity resonance frequency over the zero-phonon line (ZPL) and observing the light emitted from the cavity and collected by the MM fiber. This light is spectrally analyzed (see  Fig.~\ref{fig:linewidth}a). In order to study the emission spectrum at different temperatures, we cool the sample down to 4~K with several intermediate steps. Note, that when the cavity is at resonance with the SiV$^-$ emission, it merely acts as an efficient light collection system, hence not modifying the shape of the spectrum significantly.\\
To obtain the full ZPL spectrum, we sum over the SiV$^-$ signal emitted into the fundamental cavity mode for the whole cavity length range. The result is shown in Fig.~\ref{fig:linewidth}b for each temperature. At room temperature, we observe a broad emission peak around 738.7~nm with a linewidth of about 5~nm. At liquid He temperature, the A/B peak vanishes, indicating a strong phononic relaxation into the lower branch of the excited state, possibly due to strain in the diamond crystal~\cite{Sternschulte1994,Meesala2018,Clark1995}. The remaining C/D peak features a linewidth of $(310\,\pm \,10)$~GHz. We attribute the broader linewidth with respect to earlier findings in bulk diamond~\cite{Rogers2014b} to inhomogeneous broadening of the SiV$^-$ ensemble due to local strain from Si implantation~\cite{Evans2016} as well as mechanical stress induced from improper bonding.\\
At 60~K and 80~K, the doublet structure is fitted by a double Lorentzian peak function with equal linewidth, resulting in center positions at $(736.57\,\pm\,0.02)~$nm and $(737.25\,\pm\,0.01)~$nm (Fig.~\ref{fig:linewidth}b). The observed splitting is $\delta = (370\,\pm \,20)$ GHz, neglecting a small dependency of the temperature~\cite{Jahnke2015}. We interpret the peaks as the A/B and C/D transition in the electronic SiV$^{-}$ level (see Fig.~\ref{fig:expSetup}e). The ground state splitting into two doublets with an expected value of $\Delta_{\text{gs}}=50$ GHz, respectively, is lower than the inhomogeneous broadening of the ensemble and hence not resolvable. The measured splitting between the two peaks is larger than the combined value for both splittings $\Delta_{\text{gs}} + \Delta_{\text{es}}=310$~GHz~\cite{Hepp2014}.\\
A possible reason could be strain in the diamond membrane~\cite{Meesala2018}, due to imperfect bonding caused by the polycrystalline layer between parts of the membrane and the plane mirror~(see Fig.~\ref{fig:membraneApp}d) or due to different temperature expansion coefficients of diamond and the mirror substrate. 
The mean ZPL center position and linewidth of the SiV$^{-}$ ensemble follow the cubic dependence with decreasing temperature (Fig.~\ref{fig:linewidth}c)~\cite{Jahnke2015}. The center fit approaches $(736.86\,\pm \,0.03)$~nm for low temperatures, in good agreement with other findings~\cite{Jahnke2015,Meesala2018}, but contradicting the claim of strain present in the diamond crystal.  

\subsection{Cavity-induced enhancement of the spontaneous emission rate}\label{subSec:lifetime}

\begin{figure}
	\includegraphics[width=0.49\textwidth]{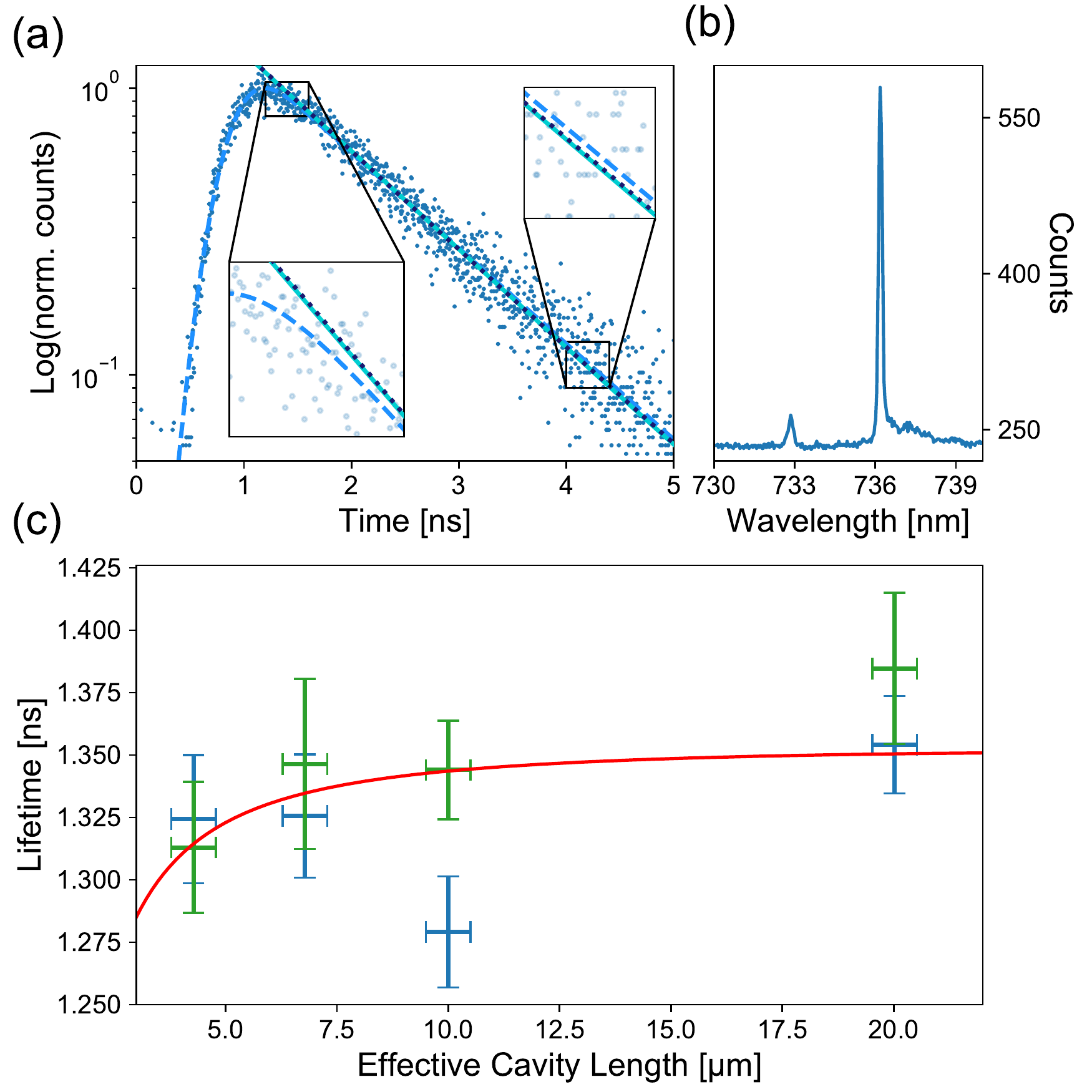}
	\caption{Excited state lifetime measurement of SiV$^{-}$ ensemble. \textbf{a} Typical time trace after pulsed excitation of the SiV$^-$ ensemble, for the cavity being resonant to the A/B transition (blue points). The corresponding fit functions are: exponentially modified Gaussian decay function (dashed), mono-exponential decay (dotted) and Kohlrausch function (solid). \textbf{b} SiV$^{-}$ transmission spectrum with cavity resonant to A/B transition and green excitation. \textbf{c} Cavity length dependent lifetimes, extracted from all three decay models fit to the time traces. Blue: cavity enhancing the A/B transition, green: cavity enhancing the C/D transition, red: fitted model following Eq.~\ref{eq:tauReduction} using $\tau_0=1.36(2)$~ns and $\eta_{\text{QE}}=0.51(35)$ (see text).}
	\label{fig:lifetime}
\end{figure}

Purcell pioneered the enhanced emission of emitters into a cavity mode and the reduction of the excited state lifetime~\cite{Purcell1946}. 
For a solid state emitter coupled to an optical microcavity, only the radiative decay into the ZPL, $\gamma_{r,\text{zpl}}=\zeta\,\gamma_{r}$, is enhanced by the cavity, with $\zeta$ being the Debye-Waller factor. The total decay rate consists of three contributions $ \gamma_{\text{tot}}=\gamma_{nr}+\gamma_{r,fs}+\gamma_{r,c} $, with the non-radiative decay $\gamma_{nr}$, the radiation into free space $\gamma_{r,fs}\approx\gamma_r$, which is approximated by the radiative decay at an absent cavity, and the radiation into the cavity mode $ \gamma_{r,c}=\zeta F_p \gamma_{r}$, which is modified by the Purcell factor $F_p$. 
This results in a reduction of the lifetime by~\cite{Benedikter2017}

\begin{equation}\label{eq:tauReduction}
	\frac{\tau_0}{\tau_c}=1+\eta_{\text{QE}}\, \zeta \, F_p
\end{equation}

with the quantum efficiency $\eta_{\text{QE}}=\gamma_r/(\gamma_r+\gamma_{nr})$. The Purcell factor is given by

\begin{equation}\label{eq:purcellFactor}
	F_p=\xi^2 \,\frac{3(\lambda/n)^3 Q_{\text{eff}}}{4\pi^2 V_m}  
\end{equation}

with $ \xi=\lvert\frac{\vec{\mu}\cdot\vec{E}}{\mu E_0}\rvert $ being the spatial and directional overlap of the dipole moment with the cavity light field mode, $ n $ the refractive index of the host material, $ Q_{\text{eff}}=(Q_{\text{em}}^{-1}+Q_{\text{c}}^{-1})^{-1} $ the effective quality factor, derived by the quality factor of the emitter ensemble $ Q_{\text{em}} $ and the cavity $ Q_{\text{c}} $~\cite{Meldrum2010}. The cavity mode volume is given by $ V_m = \frac{\pi}{4} w_0^2 L_{\text{eff}} $, with  $ w_0 $ the waist of the cavity mode and the effective cavity length $ L_{\text{eff}} $, which factors in the diamond membrane and the penetration of the light field into the dielectric mirror stacks, weighted by the local energy density of the light field mode~\cite{vanDam2018}.\\
We perform time-correlated single-photon counting after pulsed excitation of the SiV$^{-}$ ensemble to investigate the influence of the cavity on the excited state lifetime of the emitters. The ensemble is excited by a pulsed laser source\footnote{PicoQuant LDH-P-FA-530B} at 532~nm via the SM fiber, and the emission is detected via the MM fiber by an avalanche photodiode\footnote{Perkin Elmer SPCM-AQR}, after passing several spectral filters. The signal is fed to a time-correlated single photon counting module to record the time trace of the decay of the excited state. All measurements were performed at 77~K temperature in order to grant a homogeneous overlap of the ensemble with the cavity mode. Note that due to the large free spectral range of the microcavity, no significant emission of the ensemble is supported by another cavity mode.\\
Fig.~\ref{fig:lifetime}a shows the time trace of detected events after pulsed excitation for the case of the cavity enhancing the transition A/B at an effective cavity length of $L_{\text{eff}}=(10.0\, \pm \,0.5)~$\textmu m. The time traces were fit using three different models: a mono-exponential decay, a stretched exponential decay function (Kohlrausch function) $I_k(t) \propto \text{exp} [-(t/\tau_c)^\beta] $ to take potential multi-exponential decays into account~\cite{Berberan-Santos2005}, and a convolution of a Gaussian function with an exponential decay (EMG function) which includes the instrument response function.\\
The fit results of the mono-exponential and the stretched exponential decay model agree with each other, indicating that multi-exponential decay only plays a negligible role. Furthermore, the absence of bi-exponential decay implies that no collective emitter dynamics are present~\cite{Temnov2005}. The EMG model results give $ (3.6\,\pm\,0.8)$\% larger values for the excited state lifetime. We take the minimum and maximum values of the lifetime from the different fit models as a conservative estimate of the uncertainty of the measured lifetime. Note that off-resonant lifetime measurements were not possible due to parasitic background emission of the fibers leading to a low signal-to-noise ratio, caused by dopants and impurities in the glass material of the fiber cables.

The Purcell factor increases with reduced cavity length (Eq.~\ref{eq:purcellFactor}).
To observe this effect in the experiment, we measure the excited state lifetime for varying cavity lengths and set the cavity to be resonant to the A/B and C/D transition, respectively, for each cavity length. The result is shown in Fig.~\ref{fig:lifetime}c. We fit the data with a model following Eqs.~\ref{eq:tauReduction} and \ref{eq:purcellFactor}, where $Q_{\text{eff}}$ is experimentally determined, $w_0$ and $L_{\text{eff}}$ are calculated by a coupled Gaussian beams model and the transfer matrix model~\cite{Janitz2015,vanDam2018}, $\xi$ is calculated from the electrical field at the implantation depth given by the SRIM simulation, and $ \tau_0 $ and $ \eta_{\text{QE}}$ are free parameters. The measured data is shown in Fig.~\ref{fig:lifetime}c, as well as the fit model for the parameters $ \tau_0=(1.36\, \pm \, 0.02)$~ns, and $\eta_{\text{QE}}=0.51\,\pm \,0.35$.\\
The determined off-resonant lifetime of the ensemble lies in the range of about 1-4~ns of earlier results \cite{Rogers2014b,Jahnke2015,Neu2011,Neu2013,Sternschulte1994}, as well as the quantum efficiency for bulk-like diamond samples~\cite{Riedrich-Moeller2014,Zhang2018}. Note, that the fit function can only give a coarse approximation for the quantum efficiency, since it only becomes relevant for small effective cavity lengths. Using the extracted free space lifetime from the fit, we determine a maximum cavity induced lifetime reduction of $(6.8\,\pm \, 2.8)\%$ for the A/B transition and $(3.2\,\pm\,1.9)\%$ for the C/D transition. We estimate the Purcell factor using the fit parameters together with Eq.~\ref{eq:tauReduction} to be $F_{p,\text{meas}}=0.075\,\pm\,0.074$ for the C/D transition at the shortest effective cavity length, only slightly above the threshold of significance, but in agreement with the expected value from Eq.~\ref{eq:purcellFactor}: $F_{p,\text{th}}=0.071\,\pm\,0.018$.

\section{Discussion and Outlook}
In this work, we present an experimental apparatus to couple color center ensembles hosted in diamond membranes to a fiberbased microcavity with a small mode volume at temperatures down to 4~K. The setup shows reproducible performance after several thermal cyclings. The design allows for the fast exchange of samples and thus grants flexibility with respect to the color center species~\cite{Bradac2019} as well as advances in sample quality. The $\rm ^{3}He$-$\rm ^{4}He$ dilution refrigerator should enable the cooling of the sample to the~mK range, and can further be upgraded to include a vector magnet, allowing both the spectral resolving of the spin qubit and long coherence times for the SiV$^-$ qubit~\cite{Sukachev2017a,Becker2018}, encoded in the Zeeman sublevels of the ground state of the lower branch~\cite{Rogers2014a,Pingault2014}.\\
In future work, we plan to improve the fabrication and treatment of the diamond membrane in order to reduce the residual strain, which is present in the current sample. Off-resonant excitation of the SiV$ ^- $ ensemble with a wavelength closer to the zero-phonon line of the transition and the choice of pure silica core optical fiber cables can be used to decrease background fluorescence of the input and output fibers in order to to detect single emitters. Going to single emitters, the Purcell enhancement can be boosted in multiple ways: The quality factor of an SiV$^-$ emitter can be enlarged up to $ 376\times10^6 $ for the lifetime-limited linewidth of 141~MHz by cooling down the sample further. This would make the use of mirror coatings with higher reflectivity useful. Fiberbased optical resonators with finesse values in the range of $ 10^{5} $ are feasible with current technology. A more rigid design of the experimental insert~\cite{Gallego2016} and advanced locking techniques combined with higher-order low-pass filters in the electronic lock~\cite{Janitz2017} circuit should enable the stabilization of cavities with finesse values of about $ 2\times 10^{4} $~\cite{Casabone2020}. This raises the cavity quality factor to $ 4.1\times10^{5} $  for the smallest accessible effective cavity length of about 3.5~\textmu m and therefore boosts the single emitter Purcell factor to about 144 with a near-unit collection efficiency of $\beta= F_p/(1+F_p)=99.3$\%, reaching the strong coupling regime.\\
To measure the properties of color centers without the influence of the microcavity, an additional, non-coated fiber could be implemented into the setup.\\
Concerning the diamond membrane, a new generation of samples has shown reduced surface roughness values down to $ \sigma_{\text{rms}}=1~$nm, which is close to state-of-the-art roughness values reported from 0.3~nm to about 1~nm \cite{Janitz2015,Ruf2019,Riedel2020}. Using an advanced application technique, based on the controlled transfer of the membrane using a focused ion beam (FIB) device allows for a cleaner and improved bonding, effectively increasing $ Q_{\text{c}} $ with the incorporated membrane. Lower implantation doses and high temperature annealing can lead to narrow-linewidth homogeneous emitters~\cite{Evans2016,Yamamoto2013,Orwa2011,Osterkamp2019}.

\vspace{0.5cm}

\noindent \textbf{Acknowledgements:}
We thank Morgane Gandil, Philipp Fuchs, Elke Neu and Christoph Becher for helpful discussions as well as characterization measurements of the membrane samples. We thank Michael Kieschnick and Jan Meijer for the implantation of Si into the samples. We thank Lachlan J. Rogers and Fedor Jelezko for helpful discussions at the early stages of the experiment. We thank Kumaravelu Ganesan for assistance with membrane samples` fabrication.            
The work was performed in part at the Melbourne Centre for Nanofabrication~(MCN) in the Victorian Node of the Australian National Fabrication Facility~(ANFF). AN is supported by the Australian Research Council via Linkage Grant LP160101515. Experiments were partly performed using the Qudi software suite~\cite{Binder2017}. MS, YH, DH and FSK acknowledge financial support by the Bundesministerium f\"ur Bildung und Forschung via Q.Link.X. MS, YH and FSK acknowledge financial support by the VolkswagenStiftung.

\bibliographystyle{apsrev}
\bibliography{SiVmicrocav_Resub_arxiv_Lit}

\end{document}